\documentclass[twocolumn,tightenlines,prd,aps,showpacs,floatfix]{revtex4}
\usepackage{graphicx}
\usepackage{color}
\usepackage{amsmath}
\usepackage{amssymb}
\usepackage[english]{babel}

\newcommand{\fdiff}{\mathcal{D}}

\newcommand{\nn}{\nonumber}
\newcommand{\be}{\begin{eqnarray}}
\newcommand{\ee}{\end{eqnarray}}

\begin{document}

\title{The sign problem across the QCD phase transition}

\begin{abstract}
    The average phase factor of the QCD fermion determinant signals the
    strength of the QCD sign problem. We compute the average phase factor as
    a function of temperature and baryon chemical potential using 
a two-flavor NJL model. 
This allows us to study the strength of the sign problem at and
    above the chiral transition. It is discussed how the $U_A(1)$ anomaly
    affects the sign problem. Finally, we study the interplay between the 
    sign problem and the endpoint of the chiral transition.    
\end{abstract}

\author{Jens O. Andersen$^{\,a,\,b}$}
\author{Lars T. Kyllingstad$^{\,b}$} 
\author{Kim Splittorff$^{\,c}$}
\affiliation{
    $^{a}$Niels Bohr International Academy,
        Blegdamsvej 17,
        DK-2100 Copenhagen,
        Denmark\\
    $^{b}$Department of Physics,
        Norwegian University of Science and Technology,
        H\o gskoleringen 5,
        N-7491 Trondheim,
        Norway \\
    $^c$Niels Bohr Institute,
        Blegdamsvej 17, 
        DK-2100 Copenhagen,
        Denmark
}

\date{\today}


\maketitle

\section{Introduction}

Suppose that there was no sign problem in QCD at nonzero baryon chemical
potential. Then, by means of lattice QCD, it would be possible directly 
to study the phase diagram of QCD. In particular, it would be of great
interest to get a direct computation of the preferred ground state of
strongly interacting matter for low temperature and chemical potential
of order $\Lambda_\mathrm{QCD}$. This region is relevant for dense
stars, and a number of theoretically motivated ground states has been
proposed \cite{Klebanov,AndyJac,CFL,quarkyonic}---see
Ref.~\cite{RW} for a review. Clearly, a reliable lattice QCD
test of these theoretical predictions could lead to substantial
progress.
It would also be of relevance to study chemical potentials
and temperatures in the range of 100-400 MeV where 
ultrarelativistic heavy-ion collisions are conducted.

In order to show that it is the sign problem which is the main
obstacle for such first principle numerical investigations of the
QCD phase diagram,
let us for a moment consider phase-quenched QCD. In
the phase-quenched version, one replaces the fermion determinant by its
absolute value. Standard lattice techniques 
can therefore be applied. Studies
of this theory on the lattice show a rich phase diagram
\cite{KS1,KS2,deFSW}. For two flavors the phase-quenched theory is realized
by coupling the 
chemical potential to the third component of the isospin rather than
to the quark number~\cite{AKW}, and the phases obtained numerically can be
understood physically. For example, the phase quenched lattice data show that a
second-order phase transition takes place at zero temperature once the
isospin chemical potential reaches half of the pion mass
\cite{KS1,KS2,deFSW}. This is understood to be the  transition into a phase
with a Bose condensate 
of pions. Besides offering direct insights into an
alternative direction of the QCD phase diagram, the phase quenched
studies also demonstrate that without the sign problem it would be
possible also to study nonzero quark chemical potential by means of
lattice QCD.

While a direct solution of the QCD sign problem eludes us at present,
progress has been made to circumvent the sign problem
\cite{deFPh,owe1,maria,owe2,gupta,Allton1,Allton2,Allton3,Glasgow,fodor1,
fodor2,KW,FOC,AFP,aarts,Schmidt,Ejiri}. 
The success of these approaches is linked to the degree of cancellations
which take place due to the sign problem. One way to measure the degree
of such cancellations is by means of the average of the phase factor
\be
e^{2i\theta} = \frac{\det(D+\mu\gamma_0+m)}{\det(D+\mu\gamma_0+m)^*}.
\ee
If the average of the phase factor is tiny then severe cancellations
take place in the path integral defining the QCD partition function
and the validity of lattice computations should be examined carefully.     
Studies of the average phase factor are therefore important in order
to understand lattice gauge computations which circumvent the sign
problem.   

Chiral perturbation theory has been used to predict the behavior of
the average phase factor \cite{phase-long,1loop,BW}. These predictions
have been confirmed by lattice QCD measurements. Of course,
as the temperature approaches the critical one, chiral perturbation
theory is no longer valid since the more massive modes are no longer
Boltzmann suppressed. For example, it has been demonstrated in
Ref.\ \cite{Massimo} that the effect of the nucleons on the average phase
factor can be tracked in lattice QCD simulations just below the
critical temperature.

The purpose of the present study is to examine the average phase
factor also at and above the chiral phase transition, where we can not
trust chiral perturbation theory. In order to carry out such a study
we necessarily need to rely on a specific model. Here we have chosen
the two-flavor NJL model. This is perhaps the most widely used
framework to model aspects of the QCD phase diagram
\cite{Barducci:2004tt,Ratti:2004ra,He:2005nk,Ebert:2005cs,
Lawley:2005ru,Andersen:2007qv,Abuki:2008wm}.
For a review, see Ref.\ \cite{buballa}.

Since the phase factor is what separates phase quenched QCD
from full QCD, the expectation value of the phase factor in the phase
quenched theory is simply 
\be
\label{eq:phase_zratio}
\langle e^{2i\theta}\rangle = \frac{Z_{1+1}}{Z_{1+1^*}},
\ee 
where the subscript $1+1$ refers to the ordinary two-flavor theory and
$1+1^*$ refers to the two-flavor phase quenched theory (since it has a
flavor and a conjugate flavor). 

In the present paper we will evaluate this ratio within the NJL model
at mean field level. The general structure of the saddle point 
approximation gives   
\be
\label{MFgeneral}
\langle e^{2i\theta}\rangle = \frac{Z_{1+1}}{Z_{1+1^*}} =
\frac{\sqrt{\det H_{1+1^*}}}{\sqrt{\det H_{1+1}}}
e^{-\beta V (\Omega_{1+1}-\Omega_{1+1^*})},
\ee
where $\Omega$ is the mean field free energy density and $H$ is the
Hessian matrix describing the fluctuations at the saddle point. 

For small values of the chemical potential, the mean free energies in the $1+1$
theory and the $1+1^*$ theory are identical. In this case, the
prefactor---the ratio
of the Hessian determinants---determines the average phase factor. 
For $\mu>m_\pi/2$, 
the phase quenched theory enters
the Bose condensed phase mentioned
above. The two mean-field free energies are therefore different in
this phase. 

This effect was first understood within chiral perturbation
theory. It is also present when the average phase factor is evaluated
within a chiral random matrix model \cite{Han} and also results from an
exact computation in the $\epsilon$-regime of chiral
perturbation theory~\cite{SV-factorization,Juliane}.       

The $U_A(1)$ axial anomaly affects the order of the chiral phase
transition at zero chemical potential \cite{PW}. A recent
study~\cite{fukushima} shows that the location of the endpoint of the
first order chiral transition can be highly sensitive to the strength
of the $U_A(1)$ breaking. Here we compute the effects of the $U_A(1)$
breaking on the average phase factor within the two-flavor NJL model.  

Since the strength of the sign problem sets a practical upper limit for
lattice QCD it is crucial to understand the value of the average phase
factor at the endpoint. 
Here we discuss the interplay between the endpoint and the
average phase factor. Perhaps somewhat surprisingly we show that the
average phase factor grows (indicating a milder sign problem) as one 
approaches the endpoint. The reason is that the massless mode 
associated with the endpoint induces a pole in the
prefactor of Eq.~\eqref{MFgeneral}. 

The paper is organized as follows. 
In Sec.~II, we briefly discuss the two-flavor NJL model and its symmetries.
In Sec.~III, we study the average phase factor without axial symmetry
breaking, and in Sec.~IV, we investigate the effects of axial symmetry
breaking. In Sec.~V, we look at the interplay between the sign problem
and the endpoint of the chiral transition, and in Sec.~VI we
summarize our results.

\section{The NJL Model and the average phase factor}
In this section, we briefly discuss the two-flavor NJL model and our
conventions. Note in particular that we are including the prefactor
in our mean-field calculation of the partition function,
cf. Eq.~(\ref{eq:z_njl}).

The fermionic part of the two-flavor QCD Lagrangian at finite baryon
chemical potential and finite isospin chemical potential is
\begin{equation}
  \bar\psi (D - m + \mu_B B \gamma_0 + \mu_I I_3 \gamma_0) \psi,
\end{equation}
where $B = \mathrm{diag}(1/3, 1/3)$, $I_3 = \tau_3/2$, and $D$ is
the QCD fermion operator at zero chemical potentials. Since the
Lagrangian is quadratic in the fermionic fields, 
the fermions can be integrated
out exactly, yielding the fermion determinant
\be   \nonumber
  && \det(D + m - \mu_B B \gamma_0 - \mu_I I_3 \gamma_0) \\
  && = \det(D + m - \mu \gamma_0 - \delta\mu \gamma_0)   \nonumber \\
  && \phantom{=} \times \det(D + m - \mu \gamma_0 + \delta\mu \gamma_0)\;,
\label{eq:determinant}
\ee
where the chemical potentials $\mu$ and $\delta\mu$ are
\begin{equation} \label{eq:mu_def}
  \mu \equiv \frac{\mu_B}{3}
  \qquad \mathrm{and} \qquad
  \delta\mu \equiv \frac{\mu_I}{2}.
\end{equation}
$\mu$ is thus the quark chemical potential.

As mentioned in the introduction, phase-quenched QCD with two flavors
is realized by setting $\mu = 0$ and interpreting $\delta\mu$ as
the quark chemical potential instead.

The NJL Lagrangian is given by~\cite{buballa}
\begin{equation} \label{eq:vacuum_L}
  \mathcal L
  =
  \mathcal L_0 + (1-\alpha) \mathcal L_1 + \alpha \mathcal L_2,
\end{equation}
where the various terms are
\begin{eqnarray}
  \mathcal L_0 &=& 
  \bar\psi (i \gamma^\mu \partial_\mu - m) \psi
  \\
  \mathcal L_1 &=&
  G \left[
    (\bar\psi \psi)^2
    + (\bar\psi \boldsymbol\tau \psi)^2
    + (\bar\psi i \gamma^5 \psi)^2
    + (\bar\psi i \gamma^5 \boldsymbol\tau \psi)^2
    \right]\nn
  \\
&&\\ \nonumber
        \mathcal L_2 &=&
        G \left[
          (\bar\psi \psi)^2
          - (\bar\psi \boldsymbol\tau \psi)^2
          - (\bar\psi i \gamma^5 \psi)^2
          + (\bar\psi i \gamma^5 \boldsymbol\tau \psi)^2
          \right].\nn
\\ &&
\end{eqnarray}
In the two-flavor case, $\psi$ is an isospin doublet, i.e.
\begin{equation}
  \psi = \binom{u}{d},
\end{equation}
and $m=\mathrm{diag}(m_u,m_d)$ is the quark mass matrix.
In the following, we take $m_u = m_d = m_0$.
Also, $\boldsymbol\tau = (\tau_1, \tau_2, \tau_3)$, where the $\tau_i$s
are the Pauli matrices acting in isospin space.

The Lagrangian \eqref{eq:vacuum_L} is invariant under global
$SU(N_c)$ and $U(1)_B$ symmetries. The latter
corresponds to conservation of baryon number. The interaction
terms $\mathcal L_1$ and $\mathcal L_2$ both 
have a a global
$SU(N_f)_\mathrm{L} \times SU(N_f)_\mathrm{R}$,
which is broken explicitly to a $SU(N_f)_\mathrm{L+R}$ by a
nonzero value of $m_0$. Finally, $\mathcal L_1$ has a $U(1)_A$
symmetry which is not shared by $\mathcal L_2$. This means that we
can use the parameter $\alpha$ to control the strength of the axial
symmetry violation in the model, and when $\alpha=0$, 
the Lagrangian~(\ref{eq:vacuum_L}) is $U(1)_A$ symmetric.

We introduce chemical potentials for baryon number density and
isospin density by, respectively, adding the terms
\begin{eqnarray}
  \mathcal L_B &=& \mu_B \bar\psi \gamma^0 B \psi,  \\
  \mathcal L_I &=& \mu_I \bar\psi \gamma^0 I_3 \psi.
\end{eqnarray}
Using the definitions in Eq.~(\ref{eq:mu_def}), we can now write
the Lagrangian as
\begin{eqnarray}
  \mathcal L
  &=&
  \bar\psi (i \gamma^\mu \partial_\mu - m_0 
  + \mu \gamma^0 + \delta\mu \gamma^0 \tau_3) \psi \\
  && + G \left[
    (\bar\psi \psi)^2
    + (\bar\psi i \gamma^5 \boldsymbol\tau \psi)^2
    \right]
  \nonumber \\ 
  &&  + (1 - 2\alpha) G \left[
    (\bar\psi \boldsymbol\tau \psi)^2
    + (\bar\psi i \gamma^5 \psi)^2
    \right]. \nn
\end{eqnarray}
When $\alpha = 1/2$, the terms in the scalar-isovector and
pseudoscalar-isovector channels vanish, and one recovers the standard
NJL Lagrangian.

Next, we introduce four auxiliary scalar fields $\sigma$
and $a_i$, 
and four pseudscalar fields
$\eta$ and $\pi_i$, ($i=1,2,3$),
by adding the following terms to the Lagrangian:
\be\nonumber
  \mathcal L_\mathrm{aux} &=&
  - \frac{1}{4 G}\left[\left(\sigma-2G\bar{\psi}\psi\right)^2 + 
\left(\pi_i - 2 G \bar\psi i \gamma^5 \tau_i \psi\right)^2\right]
\\&& \nonumber
  - \frac{1-2\alpha}{4 G}\left[(\eta - 2 G \bar\psi i \gamma^5 \psi)^2
+\left(a_i -2G\bar\psi \tau_i \psi\right)^2
\right]\;.
\\ &&
\ee
If we use the equation of motion for the auxiliary fields to eliminate
them, we recover the original Lagrangian.
The quartic interaction terms in the Lagrangian vanish and we obtain
\begin{eqnarray}   \nonumber
  \mathcal L
        &=&
  \bar\psi \left[i \gamma^\mu \partial_\mu - m_0
    + \mu \gamma^0 + \delta\mu \gamma^0 \tau_3
    - \sigma \right. \\
  && \phantom{\bar\psi[} \left.- i \gamma^5 \tau_i \pi_i 
     - (1-2\alpha)(
    i \gamma^5 \eta
    + \tau_i a_i)
  \right] \psi
  \nonumber \\ &&
        - \frac{1}{4 G} (\sigma^2 + \pi_i \pi_i)
        - \frac{1-2\alpha}{4 G} (\eta^2 + a_i a_i).
\label{eq:njl_full_lagr}
\end{eqnarray}
Based on the last term in Eq.\ \eqref{eq:njl_full_lagr}, it is evident
that we must require $\alpha \leq 1/2$ for the
theory to be stable, which means that the $\mathrm{U}_A(1)$ symmetry is
maximally violated for $\alpha = 1/2$.

To allow for a chiral condensate $\langle \bar \psi \psi \rangle$
and a charged pion condensate
$\langle \bar\psi i\gamma^5 \tau_1 \psi \rangle$, we introduce
nonzero expectation values for the fields $\sigma$ and $\pi_1$.
Expanding around these values, we have that
\begin{eqnarray}
  \sigma &\to& \sigma
  -2 G \langle \bar \psi \psi \rangle, \\
  \pi_1 &\to& \pi_1
  -2 G \langle \bar \psi i \gamma^5 \tau_1 \psi \rangle.\nn
\end{eqnarray}
For notational simplicity, we introduce the quantities
\begin{eqnarray}
  M &\equiv& m_0 - 2 G \langle \bar \psi \psi \rangle, \\
  \rho &\equiv& - 2 G \langle \bar \psi i \gamma^5 \tau_1 \psi \rangle.\nn
\end{eqnarray}
We can now write the Lagrangian as
\begin{eqnarray}\nonumber
  \mathcal L
  &=&
  \bar\psi \left[i \gamma^\mu \partial_\mu - M
            - i \gamma^5 \tau_1 \rho
            + \mu \gamma^0 + \delta\mu \gamma^0 \tau_3
            \right. \\
            && \phantom{\bar\psi [} \left.
            - \sigma
            - i \gamma^5 \tau_i \pi_i
            - (1-2\alpha)(
            i \gamma^5 \eta
                + \tau_i a_i)
                \right] \psi
  \nonumber \\ &&
  - \frac{1}{4 G} (\sigma^2 + \pi_i \pi_i)
  - \frac{1-2\alpha}{4 G} (\eta^2 + a_i a_i)\;.
\end{eqnarray}
The pion condensate $\rho$ breaks parity and
isospin symmetry.
    
Note that when $\alpha \neq 1/2$ we should, in general, allow for
nonequal $\langle \bar u u \rangle$ and $\langle \bar d d \rangle$
condensates, since
$\langle a_3 \rangle = - 2 G (\langle \bar u u \rangle
- \langle \bar d d \rangle)$. However, in this paper we will only
consider the special cases $(\mu \neq 0, \delta\mu = 0)$ and
$(\mu = 0, \delta\mu \neq 0)$, and in both these cases
$\langle \bar u u \rangle = \langle \bar d d \rangle$.

After applying the shifts in Eq.\ \eqref{eq:shift_sigma},
the Lagrangian is quadratic in the fermion fields, which means
that they can be explicitly integrated out. Thus, the partition
function of the NJL model is
\begin{equation}
  Z_\mathrm{NJL}
  =
  \int \fdiff\sigma\prod_{i=1}^3\fdiff\pi_i \fdiff\eta\,\prod_{j=1}^3
\fdiff a_j \;
  e^{-S_\mathrm{eff}[\sigma,\pi_i,\eta,a_j]},
\end{equation}
where the effective action is given by
\be\nonumber
        S_\mathrm{eff}[\sigma,\pi_i,\eta,a_i] & = & 
  - \log \det K \\ \nonumber
&& \hspace{-25mm} + \frac{1}{4 G}  \int d^3 x \int_0^\beta d\tau \bigg[
    \sigma^2 + \pi_i \pi_i
    + (1-2\alpha) (\eta^2 + a_i a_i)
    \bigg]\;,
\\ &&
\ee
and the matrix $K$ is defined as
\be\nonumber
 K  & \equiv &   \gamma_\mu P_\mu + m_0
  - \mu \gamma_0 - \delta\mu \gamma_0 \tau_3
  + \sigma \\
 && + i \gamma_5 \tau_i \pi_i
  + (1-2\alpha)(
  i \gamma^5 \eta
  + \tau_i a_i)\;.
\ee
The matrix
$K$ has components in Dirac, color, and isospin space, as well as
spacetime, and the determinant must be taken over all these spaces.
In general this is highly nontrivial to do analytically, but it
can be simplified somewhat by defining 
a matrix $K^{\prime}$ by
\begin{equation}
  K^{\prime} \equiv K \gamma_0;.
\end{equation}
Since $\det(\gamma_0) = 1$, this matrix has the property that
$\log(\det K') = \log(\det K)$. Unlike $K$, however, $K'$ has all
$p_0$s on the diagonal, so it can be written as
\begin{equation}
  K' = p_0 + K'(p_0=0).
\end{equation}
Thus, the eigenvalues of $K'$ (and $K$) can be written as
\begin{equation}
  \lambda_i = p_0 + \varepsilon_i,
\end{equation}
where $\varepsilon_i$ are the eigenvalues of the simpler matrix
$K'(p_0 = 0)$. The fermion part of the effective action is then
\begin{equation}
  \log \det K = \sum_i \log (p_0 + \varepsilon_i).
\end{equation}
Since the original matrix $K$ is the inverse fermion
propagator, we immediately identify $\varepsilon_i$ as the
energies of the quasiparticles in the spectrum.

The values of the condensates $M$ and $\rho$ are found by
minimizing the effective potential in the $(M,\rho)$ plane,
i.e.\ by simultaneously solving the equations
\begin{equation}
  \frac{\partial S_\mathrm{eff}}{\partial M} = 0,
  \qquad
  \frac{\partial S_\mathrm{eff}}{\partial \rho}  = 0,
\end{equation}
while ignoring the fluctuations in the fields.

Using the saddle-point approximation to evaluate the partition function
at this minimum, we find that
\begin{equation} \label{eq:z_njl}
  Z_\mathrm{NJL}
  =
  \frac{1}{\sqrt{\det H}}
  e^{-\beta V \Omega},
\end{equation}
where the thermodynamic potential $\Omega$ is the effective action
evaluated at the saddle point and divided by the four-volume $\beta V$.
$H$ is the Hessian matrix of $S_\mathrm{eff}$, defined as
\begin{equation}
  H
  =
  \left(\begin{array}{cccc}
    \frac{\partial^2 S_\mathrm{eff}}{\partial\sigma^2}
    &
    \frac{\partial^2 S_\mathrm{eff}}{\partial\sigma \partial\pi_1}
    &
    \frac{\partial^2 S_\mathrm{eff}}{\partial\sigma \partial\pi_2}
    &
    \frac{\partial^2 S_\mathrm{eff}}{\partial\sigma \partial\pi_3}
    \\
    \frac{\partial^2 S_\mathrm{eff}}{\partial\pi_1 \partial\sigma}
    &
    \frac{\partial^2 S_\mathrm{eff}}{\partial\pi_1^2}
    &
    \frac{\partial^2 S_\mathrm{eff}}{\partial\pi_1 \partial\pi_2}
    &
    \frac{\partial^2 S_\mathrm{eff}}{\partial\pi_1 \partial\pi_3}
    \\
    \frac{\partial^2 S_\mathrm{eff}}{\partial\pi_2 \partial\sigma}
    &
    \frac{\partial^2 S_\mathrm{eff}}{\partial\pi_2 \partial\pi_1}
    &
    \frac{\partial^2 S_\mathrm{eff}}{\partial\pi_2^2}
    &
    \frac{\partial^2 S_\mathrm{eff}}{\partial\pi_2 \partial\pi_3}
    \\
    \frac{\partial^2 S_\mathrm{eff}}{\partial\pi_3 \partial\sigma}
    &
    \frac{\partial^2 S_\mathrm{eff}}{\partial\pi_3 \partial\pi_1}
    &
    \frac{\partial^2 S_\mathrm{eff}}{\partial\pi_3 \partial\pi_2}
    &
    \frac{\partial^2 S_\mathrm{eff}}{\partial\pi_3^2}
  \end{array}\right),
\end{equation}
and also evaluated at the saddle point. This matrix is a measure of the
curvature of the potential at the saddle point, i.e.\ the magnitude of
the fluctuations of the fields around this point, and it is thus
related to the masses of the effective excitations.

Using the NJL model partition function, we can now write
Eq.\ \eqref{eq:phase_zratio} as
\begin{equation}\label{FexpDOmega}
  \langle e^{2 i \theta} \rangle
  =
  F(\mu_q,T) e^{-\beta V \Delta\Omega(\mu_q,T)}
\end{equation}
where the prefactor $F$ is the square root of the ratio between the
Hessian determinants in the $1+1^*$ and the $1+1$ theories, i.e.
\begin{equation}
  F(\mu_q,T)
  \equiv
  \sqrt{\frac{\det H(\mu=0,\delta\mu=\mu_q)}
    {\det H(\mu=\mu_q,\delta\mu=0)}},
\end{equation}
while $\Delta\Omega$ is the difference in thermodynamic potential
between the two cases:
\begin{equation}
  \Delta\Omega(\mu_q,T)
  \equiv
  \Omega(\mu=\mu_q,\delta\mu=0) - \Omega(\mu=0,\delta\mu=\mu_q).
\end{equation}
Here, $\mu_q$ is used to denote the quark chemical potential, so that
$\mu_q = \mu_B/3$.

In the numerical analysis, all momentum integrals have been regulated
using a sharp ultraviolet cutoff of $\Lambda=651$ MeV. We have
taken the bare quark mass and the coupling constant to be $m_0 = 5.5$
MeV and $G = 2.12/\Lambda^2$, respectively, and set the number of
colors to $N_c = 3$. With this parameter set, the model reproduces
the pion mass $m_\pi = 139$ MeV, the pion decay constant $f_\pi = 94$
MeV, and a chiral condensate of
$\langle \bar u u \rangle = - (250\mathrm{\ MeV})^3$ in the vacuum.

In the plots of the average phase factor $\langle e^{2 i \theta} \rangle$
we have set the three-volume $V$ equal to $L^3$, where $L = 4/T$. Thus,
the full four-volume is
\begin{equation}
    \beta V = \frac{64}{T^4}.
\end{equation}
This choice is inspired by lattice QCD, and corresponds to a lattice
size of $4\times 16^3$.

\section{The importance of the prefactor}
\begin{figure}[htb]
  \center
  \includegraphics{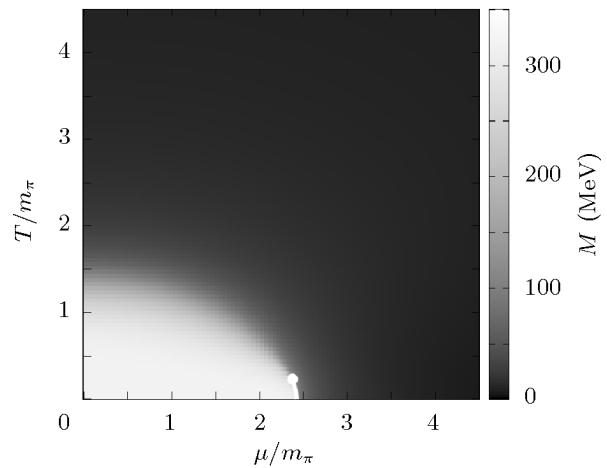}
  \caption{\label{fig:chiralB}The chiral condensate $M$ in the $1+1$
    theory  with $\alpha=0$
as a function of $T/m_{\pi}$ and $\mu/m_{\pi}$.
    The curve marks the first order phase transition, while the
    dot marks the critical point.}
\end{figure}

\begin{figure}[htb]
  \center
  \includegraphics{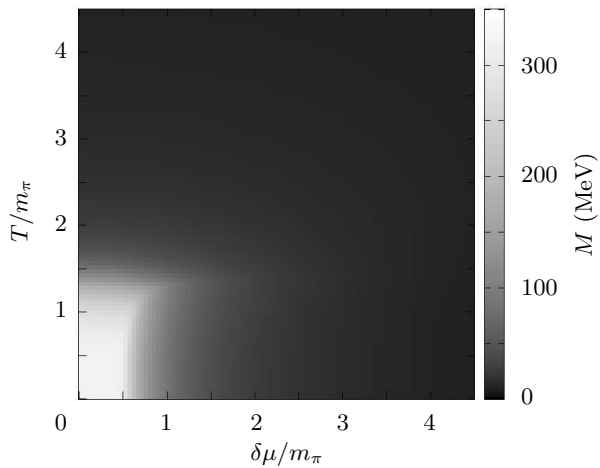}
  \caption{\label{fig:chiralI}The chiral condensate $M$ in the $1+1^*$
as a function of $T/m_{\pi}$ and $\delta\mu/m_{\pi}$
   theory with $\alpha=0$.
    }
\vspace{0.3cm}
\end{figure}

\begin{figure}[htb]
  \center
  \includegraphics{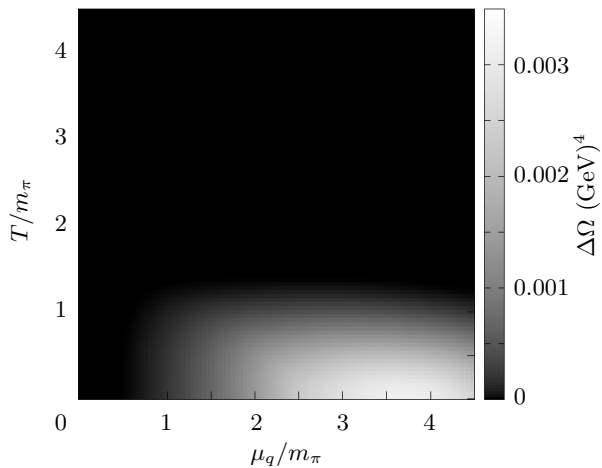}
  \caption{\label{fig:deltaomega}The difference in the free energy densities
($\Delta\Omega$),
    between the $1+1$ theory and the $1+1^*$ theory 
as a function of $T/m_{\pi}$ and $\mu/m_{\pi}$
    with $\alpha=0$.}
\end{figure}

\begin{figure}[htb]
  \center
  \includegraphics{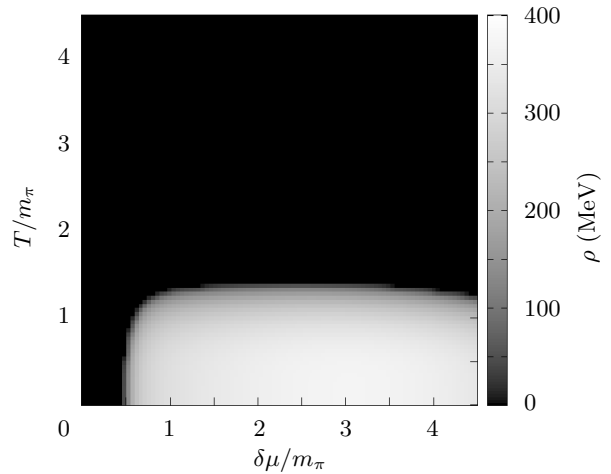}
  \caption{\label{fig:pionI}The pion condensate $\rho$ in the $1+1^*$
    theory as a function of $T/m_{\pi}$ and $\delta\mu/m_{\pi}$.
with $\alpha=0$.}
\end{figure}

In Fig.~\ref{fig:chiralB}. 
we show the chiral condensate as a function of 
function of $T/m_{\pi}$ and $\mu/m_{\pi}$.
In Fig.~\ref{fig:chiralI} we show the chiral condensate as a
function of $T/m_{\pi}$ and $\mu/m_{\pi}$.
Comparing the two, we immediately see that the 
free energies must differ for $\mu,\delta\mu>m_\pi/2$ and low $T$.
Indeed this is confirmed by computing directly the difference 
of the two free energies. This is shown in Fig.~\ref{fig:deltaomega}.
This difference is
due to the formation of a pion condensate, $\rho$ in the 
phase-quenched theory. 
In Fig.~\ref{fig:pionI} we show a plot of the pion
condensate as a function of $T/m_{\pi}$ and $\delta\mu/m_{\pi}$.
As expected, the pion condensation sets
in at $\delta\mu=m_\pi/2$ for $T=0$. As the temperature is
increased it takes a somewhat larger chemical potential to trigger the
formation of the pion condensate. As the pion condensate forms, the
chiral condensate starts dropping to zero. 

We conclude from this that  the average phase factor is exponentially
small when there is a pion condensate in the phase-quenched theory,
cf.\ Eq.\ \eqref{FexpDOmega}. However, when the pion condensate
vanishes in the phase-quenched theory, the mean field NJL free energy
is equal to that for the ordinary theory, i.e.
$\Delta\Omega(\mu_q,T)=0$ if $\rho=0$.
The average phase factor, Eq.~\eqref{FexpDOmega}, 
is then dominated by the prefactor $F$ 
throughout an extended region of
the $(\mu_q,T)$ plane. 

\begin{figure}
  \center
  \includegraphics{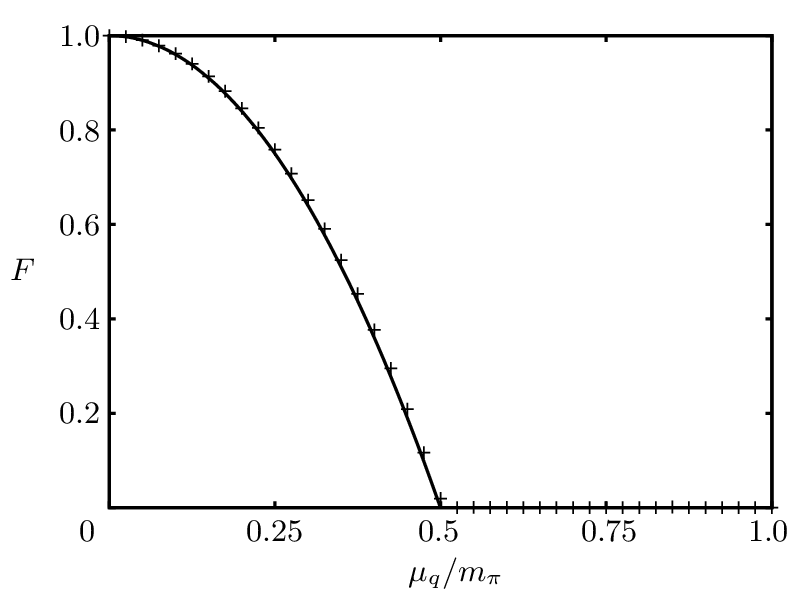}
  \caption{\label{fig:prefac_al0}
The prefactor $F$ for $T=1$ MeV and $\alpha=0$ (crosses),
    and the prediction $1-(2\mu_q/m_\pi)^2$ from Ref.~\cite{phase-long} (line).
Note that the choice of $T=1$ MeV instead of $T=0$ is purely for
computational convenience. } 
\end{figure}

The Hessian matrix measures the curvature of the potential at the
minimum and can hence be interpreted as a product of 
quasiparticle masses. 
For low temperatures where the pions dominate the partition function, 
the prefactor is simply~\cite{phase-long}
\begin{equation}
    F = 1 - \left(\frac{2 \mu_q}{m_\pi}\right)^2.
\end{equation}
This low-temperature behavior of the prefactor is confirmed by the NJL 
model computation:
In Fig.~\ref{fig:prefac_al0} we show the prefactor $F$ as predicted
in chiral perturbation theory (solid curve) and the prediction by the NJL 
model (crosses).

\begin{figure*}
  \center
  \includegraphics{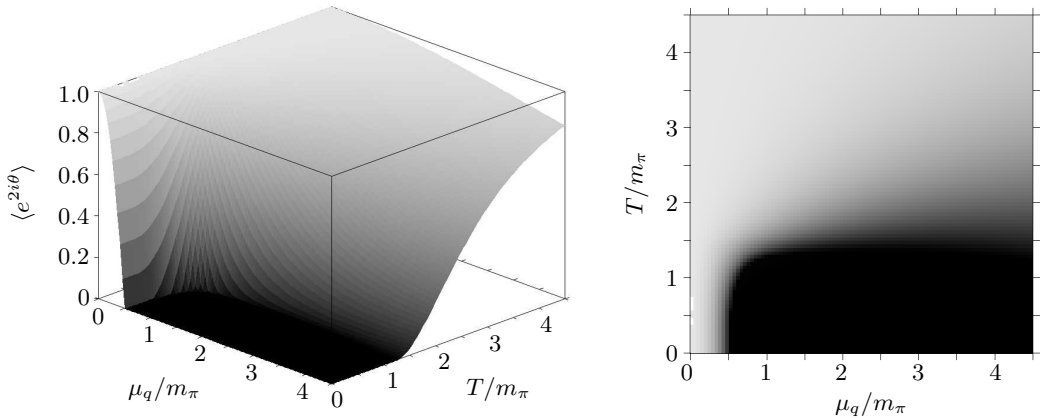}
  \caption{\label{fig:phasefac}The average phase factor
    $\langle e^{2i\theta} \rangle$
as a function of $\mu_qm_{\pi}$ and $T/m_{\pi}$ for 
$\alpha=0$.}
\end{figure*}

The full plot of the average phase factor in the
$(\mu_q,T)$ plane is given in Fig.\ \ref{fig:phasefac}. We see that
indeed the sign problem is exponentially bad in a region corresponding
to the region of pion condensation in the phase quenched
theory. Before reaching this the prefactor dominates and the average
phase factor drops smoothly from one to zero. 

The full dominance of the prefactor outside the pion-condensed region
is an artifact of the mean-field approximation. This has been
demonstrated for low temperatures in chiral perturbation theory, where
one-loop pion effects give rise to a nonzero value of $\Delta \Omega$
also where the pion condensate vanishes~\cite{phase-long}.
By comparison to
lattice QCD measurements  of the phase factor, it is seen
that both the 
prefactor and the one-loop contribution to $\Delta \Omega$ are
relevant for the standard lattice volumes used~\cite{1loop}.     
Nevertheless, the overall trend for the average phase factor also
above the critical temperature
is in agreement
with lattice studies~\cite{gattringer}.

\section{The effect of $U_A(1)$ symmetry breaking}
Instanton effects are responsible for the 
$U_A(1)$ axial anomaly in QCD and is 
known~\cite{PW} to have a
strong influence on order of the phase transition at zero chemical
potential. In Ref.\ \cite{fukushima} it was demonstrated that the strength of
the $U_A(1)$ axial anomaly is also important for nonzero $\mu$ in the
three-flavor NJL model. Here we consider the effects of the anomaly on the
phase factor within the two-flavor NJL model. As in Ref.\ \cite{fukushima} we
work with a $\mu_q$-dependent $\alpha$,
\begin{equation}
    \alpha = \frac{1}{2} e^{-\mu_q^2/\mu_0^2},
\label{inst}
\end{equation}
so that the
$U_A(1)$ symmetry is restored at $\mu_q^2\gg\mu_0^2$. 
The scale $\mu_0$
introduces a new parameter in the NJL model.
The form (\ref{inst}) is motivated by the
Gaussian suppression of the instanton density due to Debye
screening at large values of the chemical potential.

The effects of $\alpha$ on the phase diagram of the two-flavor NJL
model at nonzero $\mu$ and $\delta\mu$ have been discussed in
Ref.~\cite{buballa}. There, the focus was to demonstrate that the splitting
of the $\langle \bar{u}u\rangle$ and $\langle\bar{d}d\rangle$
transitions induced by
$\delta\mu$ is wiped out by the breaking of $U_A(1)$ symmetry.
Here the setup is somewhat
different since we do not have $\mu$ and $\delta\mu$ nonzero at the same
time; recall that we are 
looking at the ratio $Z_{1+1}/Z_{1+1^*}$. This implies that 
$\langle \bar{u}u\rangle=\langle\bar{d}d\rangle$. Thus there is no 
effects of $\alpha$ in the free energies of the mean-field two-flavor NJL 
model considered here, and the dependence of the average phase factor on
$\alpha$ is entirely due to the $\alpha$-dependence of the prefactor $F$.

\begin{figure}
  \center
  \includegraphics{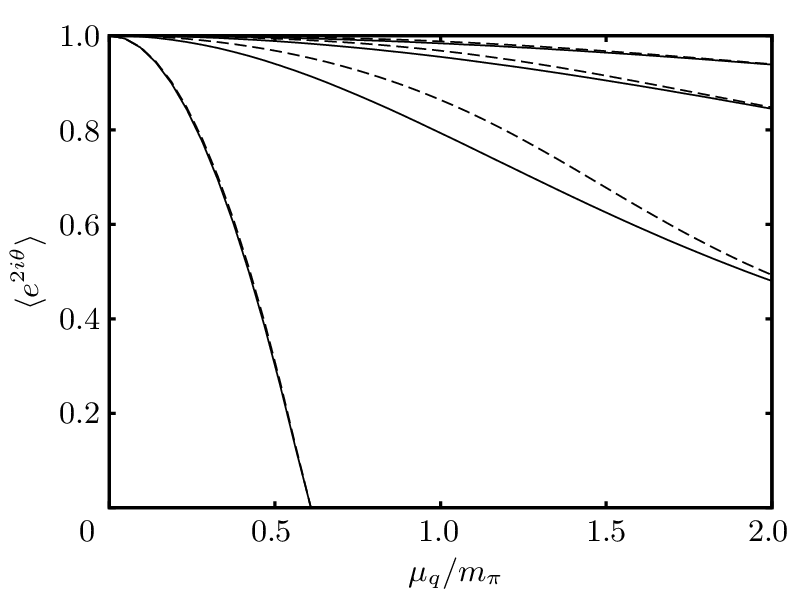}
  \caption{\label{fig:U1}The average phase factor
    $\langle e^{2i\theta} \rangle$ for $T=m_\pi$ (lower), $T=2m_\pi$,
    $T=3m_\pi$, and $T=4m_\pi$ (upper), when $\alpha=0$ (solid curves)
    and $\alpha=\tfrac{1}{2}e^{-\mu_q^2/m_\pi^2}$ (dashed curves).   }
\end{figure}

In Fig.~\ref{fig:U1}, we have plotted
the average phase factor
for four different temperatures, both with and without axial symmetry
breaking.
For the plot we have chosen $\mu_0=m_\pi$. We observe that there is no
effect of $\alpha$ for low temperature whereas the $U_A(1)$ breaking
increases the average phase factor slightly for higher temperatures. 

From a Ginzburg-Landau perspective, it is easy to see that the effect
of $U_A(1)$ breaking on the phase diagram in the two-flavor theory is 
small, since the $U_A(1)$ breaking only gives a quadratic term
in the order parameter. This
term is already present at $\alpha = 1/2$. 
For $N_f = 3$ the $U_A(1)$ breaking induces a cubic term in the
Ginzburg-Landau free energy---see for example Ref.~\cite{fukushima}. 
This term potentially has a much larger effect on the position of
the endpoint. Indeed this is what was observed for the three-flavor
NJL model in Ref.~\cite{fukushima}.

\section{The endpoint and the sign problem}

\begin{figure*}
  \center
  \includegraphics{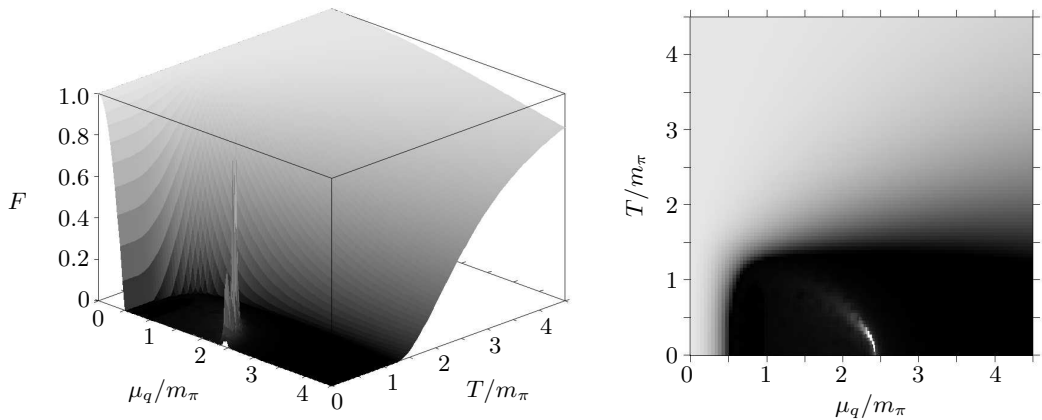}
  \caption{\label{fig:sigma-spike}
{The prefactor $F$
as a function of $\mu_q/m_{\pi}$ and $T/m_{\pi}$} for $\alpha=0$. Note
the spike at the critical point, which extends far above the
maximum value of $F$.}
\end{figure*}

\begin{figure}[htb]
    \center
    \includegraphics{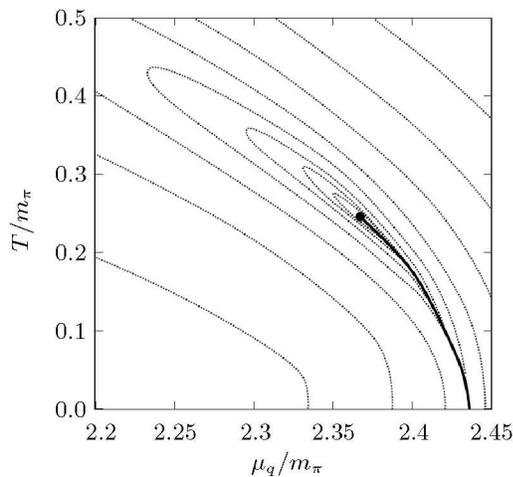}
    \caption{Contour plot of the prefactor $F$ around the critical
        endpoint, with $\alpha=0$. The solid line is the first-order phase 
transition,
        and the dot marks the endpoint. The dotted lines are contours
        of equal $F$, each separated by a power of $e$.}
    \label{fig:prefac_critical}
\end{figure}

The endpoint of the first order chiral phase transition has been 
studied
intensively in recent years. 
One conclusion which has emerged is that the
location of the endpoint in the $(\mu_q,T)$ plane is extremely sensitive
to the value of the quark masses and the fate of the $U_A(1)$
breaking. In order for lattice QCD to be able to reach the location of
the endpoint in the $(\mu_q,T)$ plane, before the numerical noise due to the
sign problem wipes out the signal, the endpoint must be located
outside the region where the average phase factor is
exponentially suppressed. 
Whether or not this is the case is unclear. Both
the position of the endpoint as well as the region where the average
phase factor is exponentially suppressed depend on the quark masses. 
In chiral Random Matrix Models \cite{Klein,Han} and in the two-flavor NJL
model, 
the location of the endpoint is always inside the exponentially
suppressed region. The endpoints found by multi-parameter reweighting
techniques in both Refs.~\cite{fodor1} and~\cite{fodor2} are located where
the exponential suppression sets in \cite{Lattice2006}. Here we
discuss how the sign problem is correlated with the critical endpoint. 

If the endpoint is located in the region where the average phase factor
is exponentially suppressed it is not accible by current lattice methods.
If, however, the endpoint is located at small values
of $\mu_q$, i.e.\ $\mu_{\rm endpoint}<m_\pi/2$, then the story is somewhat
different. Two scenarios are possible depending on what happens in the
phase-quenched theory:

{\sl 1)} The phase-quenched theory has an endpoint at exactly the same
location as the full theory. In this case
the average phase factor will not show any particular sign of the
endpoint.

{\sl 2)} The endpoint in the phase-quenched theory is not at
the same location as in the full theory. In this case the average
phase factor can grow as one approaches the critical endpoint. The
reason is simple: just like the massless mode in the pion-condensed
theory drives the prefactor $F$ to zero, see. Fig.~\ref{fig:prefac_al0},
the massless $\sigma$ at the endpoint will induce a pole
in the prefactor since the Hessian of the 1+1 theory is in the
denominator, cf.\ Eq.~\eqref{MFgeneral}. We can illustrate this by looking
at the prefactor, $F$, in the two-flavor NJL model.
In Fig.~\ref{fig:sigma-spike}, we have shown the prefactor $F$ as a function
of $\mu_q/m_{\pi}$ and $T/m_{\pi}$.
We see that at the endpoint in the $1+1$ theory 
the massless $\sigma$ has produced a peak in the exponential
prefactor. In this case the peak is suppressed by the exponential factor
even with a small volume. If, however, the endpoint is located outside
the exponentially suppressed region, the peak due to the massless
$\sigma$ may survive to much larger values of $V$. This is because the free 
energy
difference $\Delta\Omega$ can be much smaller here, since a cusp in
$\Omega_{1+1}$ does not necessarily induce a large $\Delta\Omega$. 
Of course, it must always be true that 
$0\leq\langle e^{2i\theta}\rangle\leq1$, so the pole and the
exponential factor must always be considered together to
ensure this.

As mentioned above it is unfortunately not possible to fully illustrate this
most interesting scenario in the two-flavor NJL model used here. However,
it is quite possible that it could be realized in the three-flavor
NJL model with $U_A(1)$ axial symmetry breaking.

One of the characteristic features of the critical point is the divergence
of the correlation length $\xi$.
Since the correlation length is inversely proportional to the sigma mass,
while the Hessian matrix is roughly proportional to the squares of the
masses, we expect that as we approach the endpoint,
\begin{equation} \label{eq:prefac_corrlen}
    F \sim \xi.
\end{equation}
In Fig.\ \ref{fig:prefac_critical} we have shown a contour plot
of the prefactor $F$ in
the region immediately surrounding the endpoint. A comparison of this
figure to Fig.\ 3 of Ref.\ \cite{Stephanov:2008qz}, which is a similar
plot of the correlation length $\xi$, seems to confirm 
Eq.\ \eqref{eq:prefac_corrlen}.
A notable feature of this plot is the fact that the prefactor---specifically,
the contribution from the prefactor to the $Z_{1+1}$ partition
function---is discontinuous across the phase transition.
This seems to indicate that there is a zeroth-order phase transition in the
theory. Fortunately, this discontinuity is entirely due to corrections to
the partition function of order $1/V$, so there is no inconsistency in the
infinite-volume limit.

\section{Conclusions}

We have presented the first study of the average phase factor of the
complex QCD fermion determinant within a two-flavor
NJL model. This study extends
previous studies in several directions: We have considered
temperatures not accessible to chiral perturbation theory, we have
considered the effects of the $U_A(1)$ anomaly on the sign problem 
and we have discussed the interplay between the sign problem and the
critical endpoint of the chiral phase transition. The NJL
model certainly does not capture all effects of QCD and one should not
take the precise numerical values given here as firm
predictions. However, several general conclusions can be drawn from
this study of the NJL model.

First, we have seen that the
exponential prefactor in the mean-field approximation dominates for a
large range of $\mu_q$ when the temperature is above the critical one.
This is in complete
agreement with the result obtained from a mean field random
matrix approach~\cite{Han}. For lattice QCD, this means that one
should expect that the masses of the effective excitations have a
large effect on the strength of the sign problem.    

Second, the average phase factor depends on
the $U_A(1)$ axial anomaly when the temperature is close to or above
the critical one. The present study indicates that the sign problem
may become milder when $U_A(1)$ symmetry breaking is taken into account. 

Third, we have discussed how the massless mode at the endpoint can
drive the average phase factor to larger values. Since this indicates
a milder sign problem at the endpoint, this observation is highly
relevant for lattice investigations of the endpoint. Of course the
free energies in 
the phase-quenched and the full theory must be slightly different if
the latter is to have an endpoint while the former does not. However, this
difference need not be numerically large (as it is in the pion condensed
region) and the exponential suppression of the prefactor thus may
only set in at large values of $V$. Unfortunately is it not possible
to study this effect fully within the two-flavor NJL model.

The order of the QCD phase transition depends on the 
number of flavors as well their masses~\cite{Stephanov:2007}.
It would be therefore very interesting to extend this study to 
the three-flavor case where it also known that
the effects of the $U_A(1)$ axial anomaly are greater.

\section*{Acknowledgments}
J.~O.~A.\ and L.~T.~K.\ would like to thank the
Niels Bohr International Academy, where this work was initiated, for kind 
hospitality.
K.~S.\ would like to thank Thomas Sch\"afer for discussions.
This work was supported by the Danish Natural Science Research Council
(K.~S.).

\end{document}